\documentclass[aps,showpacs,prl,twocolumn]{revtex4}
\usepackage{graphicx}
\usepackage{amsmath,amsfonts,amssymb,amsthm,amstext,amscd,eucal}
\usepackage{bbold}
\usepackage{array}
\usepackage[latin1]{inputenc}

\newcommand{\eO}{\mathcal{O}}
\newcommand{\CV}{{\cal V}}

\newcommand{\bbibitem}[1]{\bibitem{#1}}

\def\Label#1{\label{#1}%
  \smash{\hbox to0pt{\raise1ex\hbox{\tiny[#1]}\hss}}}
\def\noLabels{\let\Label=\label}
\def\nobbibitem{\let\bbibitem=\bibitem}
\def\fun#1#2{\lower3.6pt\vbox{\baselineskip0pt\lineskip.9pt
\ialign{$\mathsurround=0pt#1\hfil##\hfil$\crcr#2\crcr\sim\crcr}}}
\def\lap{\mathrel{\mathpalette\fun <}}

\begin{document}

\title[Topology from Cosmology]{Using Cosmology to constrain the Topology of  Hidden Dimensions}

\author{Joan Sim\'on$^1$, Raul Jimenez$^{1,2}$, Licia Verde$^1$, Per Berglund$^3$ and Vijay Balasubramanian$^1$}
\affiliation{$^1$Department of Physics and Astronomy, University of Pennsylvania, Philadelphia, PA-19104, USA. \\
$^2$The Observatories of the Carnegie Institution, 813 Santa Barbara St., Pasadena, CA 91101, USA.\\
$^3$Department of Physics, University of New Hampshire, Durham, NH 03824, USA.}


\date{\today}

\begin{abstract}

A four-dimensional universe, arising from a flux compactification of
Type IIB string theory, contains scalar fields with a potential
determined by topological and geometric parameters of the
internal --hidden-- dimensions.  We show that inflation can be realized
via rolling towards the large internal volume minima that are generic
in these scenarios, and we give explicit formulae relating the
microscopic parameters (e.g., the Euler number of the internal space)
to the cosmological observables (e.g., the spectral index). We find 
that the tensor-to-scalar ratio, the running of the
spectral index, and the potential energy density at the minimum are
related by consistency relations and are exponentially small in the
number of e-foldings.  Further, requiring that these models arise as
low-energy limits of string theory eliminates most of them, even if
they are phenomenologically valid. In this context, this approach
provides a strategy for systematically falsifying stringy inflation
models.
\end{abstract}


\maketitle

{\em Introduction.}---A central question in high-energy physics is
whether top-down models such as string theory can be tested or at
least constrained by experimental data.  Traditionally, we look at
particle spectra and interactions as measured in accelerator
experiments.  Here we ask whether the precision astrophysical data
that is becoming available can play a similar role.

In string theory, the four observable dimensions of the world are
usually accompanied by an additional six dimensional compact space,
whose geometry is described by a set of scalar fields.  The potentials
for these scalar fields, known as ``moduli'', are parametrized by
topological and geometric quantities derived from the structure of the
internal space, i.e. the hidden dimensions.  In a scenario where
inflation \cite{inflation} is realized by rolling moduli scalar
fields, the primordial cosmological perturbations as observed e.g, in
the properties of Cosmic Microwave Background radiation (CMB) and of
the large-scale structures, will be affected by the shape of the
potential, and hence by the
topology and geometry of the
hidden dimensions.  Progress towards realizing inflation in string
theory includes \cite{stringinflation,racetrack1,conquevinf}, for a
review see e.g., \cite{Quevedo02}.

Here, we study four--dimensional universes that arise from Type IIB
string theory \cite{GKP,KKLT}.  As usual, the internal six--dimensional 
space is chosen as a Calabi--Yau manifold, in order to
solve the equations of motion of string theory while achieving
realistic particle physics.  Generically in such settings, the scalar
potential has a minimum at a point where the internal space has a
volume that is much larger than the Planck scale
\cite{VijayPerQuevedo}.  In these models, particle physics and
inflation has been discussed in, e.g., \cite{CQS,conquevinf}.
Here, working with a simple two--scalar example, we show that inflation
can be realized as the scalar fields roll towards the large--volume
minimum.  We derive explicit formulae for the topological and
geometric parameters of the internal manifold in terms of the
observables.  These models predict specific consistency relations
between the cosmological observables, and are thus falsifiable. In
particular the tensor-to-scalar ratio, the running of the spectral
index and the cosmological constant value are related and
exponentially small in the number of e-foldings.  Requiring that these
scenarios arise as low--energy limits of string theory eliminates many
phenomenologically viable models. This provides a strategy  for systematically falsifying stringy inflation models of this kind.  Our methods and conclusions
generalize easily from the example given here to  more complex compactifications \cite{Simon+06}.

{\em Inflation in large-volume Calabi-Yau string
compactifications}.---In designing a string model, one chooses the
topology of the internal compact manifold and the strengths of the
magnetic fluxes threading it.  The shapes and sizes of surfaces within
the internal space are described by four-dimensional scalar fields, the moduli.
As shown in \cite{GKP,KKLT}, the scalar fields describing the shape of the hidden dimensions can be treated as constants for our purposes.
Likewise, the dilaton, a
special scalar field whose expectation value sets the string
coupling constant $g_s$, is also fixed to a constant.  By contrast,
the scalar fields describing the sizes of the hidden dimension, $\{\tau_i\}$, acquire a potential from quantum
effects in string theory \cite{GKP,KKLT,VijayPerQuevedo} and could
roll to realize inflation \cite{conquevinf}.  While inflation can also arise
from rolling of the axion partners of $\tau_i$, as studied in
\cite{racetrack1},  here we integrate these axions out of the
effective potential, and study rolling in $\tau_i$.

In what follows, we rely on the methodology and some of the results of
\cite{Simon+06}.  The full potential given in \cite{VijayPerQuevedo}
simplifies when the sizes of internal surfaces, as parameterized by
$\tau_i$, are large compared to the string scale.  In this case, there
is a systematic expansion in powers and exponentials of these fields
\cite{Simon+06}.  For a generic choice of internal manifold and
magnetic fluxes in it, the scalar potential is minimized at a point
where the overall volume is large and there is a hierarchy $\tau_1 \gg
\tau_2,\cdots \tau_N \gg 1$ \cite{VijayPerQuevedo}.  Thus we will work
in the large--volume approximation of the full potential.  To
facilitate explicit analysis we will study a simple model with two
scalar fields $\tau_1,\tau_2$, in terms of which the volume of the
internal space is
\begin{equation}
\CV=\mu \tau_1^{3/2}-\lambda \tau_2^{3/2} \equiv \frac{1}{x}(1-\lambda
x \tau^{3/2}) \, ; x = \frac{1}{\mu\,\tau_1^{3/2}}\,.
\end{equation}
Here $\lambda$ and $\mu$ are related to topological quantities (``intersection numbers'') of the internal manifold.

In this large--volume limit, the scalar potential given in \cite{KKLT,BBHL,VijayPerQuevedo} becomes \cite{Simon+06},
\begin{eqnarray}
   V_{\text{tot}} &=& V_{\text{up}} + V_2 + V_3 + V_4 \nonumber \\
   V_{\text{up}} &=& \gamma\,x^2 \label{potup} \\
   V_2 &=&\! -\frac{|W_0|^2g_s^4M_P^4}{4 \pi}\,2(\frac{a\,\tau}{g_s})\,\frac{A\,e^{-a\tau/g_s}}{|W_0|}\,x^2 \\
   V_3 &=&\!\frac{|W_0|^2g_s^4 M_P^4}{4 \pi}\frac{4 (\frac{a\tau}{g_s})^2}{3}\left(\frac{A\,e^{-\frac{a\tau}{g_s}}}{|W_0|}\right)^2\!\!\frac{x}{\lambda\tau^{3/2}} 
   \\
   V_4 &=& \frac{|W_0|^2g_s^4 M_P^4}{4 \pi}\,\frac{3\xi\,x }{8}\,x^2 \label{pot4}\,.
\end{eqnarray}
where, as natural in the large--volume regime, we have  assumed
\begin{equation}
  \lambda\,x\,\tau^{3/2}\ll 1\,, \quad \xi\,x\ll 1\,, \quad A\,e^{-\frac{a\tau}{g_s}}/|W_0| \ll 1\,;
 \label{eq:approx}
\end{equation}
and, for  string theory to be weakly coupled, $g_s<1$. The parameter $\gamma$ in the uplift term, $V_{\text{up}}$, measures the
strength of supersymmetry breaking effects \cite{KKLT,BKQ}. The classical superpotential, $|W_0|$,
has its origin in the magnetic fluxes in the internal space, and results
in the fixing of the shape scalars and the string coupling; in string theory $W_0$ can be
finely tuned;
 $M_P = 1 / \sqrt{8\pi G} = 2.4 \times
10^{18} {\rm GeV}$ is the reduced Planck mass. $A$ and $a$ arise
quantum mechanically in string theory, these are computable and fixed
in a given model, but below we will allow them to be free parameters.
Finally, $\xi$ is proportional to the Euler number $\chi$ of the
internal space:
$\xi = - \frac{\zeta\left(3\right) \, \chi(M)}{2(2 \pi)^3} \,, $
with $\zeta(3)\simeq 0.909$ the Riemann zeta function at 3.  The minimum of this potential has $\tau_1 \gg \tau_2$ \cite{VijayPerQuevedo}, and hence $\CV \approx 1/x$. 
In other words, our analysis is valid in the large region of the seven
dimensional parameter space $\{\gamma, \lambda,\xi,A,a,g_s,W_0\}$ and
the two dimensional field space $\{\tau, x\}$ where (\ref{eq:approx})
applies.

In general, given an arbitrary initial condition, both scalar fields
could roll simultaneously.  For simplicity, we will study conditions
under which it is a good approximation to ignore rolling in one of the
field directions.  Ref. \cite{Simon+06} shows that in the two-field
case the potential along the $\tau$ direction is not flat enough to
yield inflation. This will be important as a mechanism to end
inflation in a manner consistent with observational constraints. Thus
we search for slow roll inflationary conditions in $x$ in a region of
the potential in which any motion in $\tau$ is negligible. One
sufficient condition to restrict rolling to the $x$ direction only is
$\partial V/\partial \tau = 0$; this can be shown from the magnitude
hierarchy between the two fields. Thus, at the start of inflation, the
initial condition $(x_0,\tau_0)$ satisfies
\begin{equation}
\frac{A e^{-y_0}}{|W_0|} = 3 \lambda x_0 \tau_0^{3/2} \frac{1-y_0}{y_0(1-4y_0)}\,; ~~~ y = a\tau/g_s
\end{equation}

The scalar potential must have a local minimum towards which the
inflating field evolves and where reheating can occur. Ref. \cite{Simon+06} show that at such a minimum $y > 1$.

 {\em Slow Roll Inflation in the $x$ direction}.---To compute the slow roll parameters (e.g.,  \cite{LiddleLyth92})
\begin{equation}
\epsilon_{\rm V}\equiv\frac{M^2_{\rm P}}{2} \left( \frac{V'}{V} \right)^2\!;\, \eta_{\rm V} \equiv M^2_{\rm P} \left( \frac{V''}{V} \right)^2\!;\, \xi_{\rm V} \equiv  M^4_{\rm P}\frac{V'V'''}{V^2} \, ,
\end{equation}
we must take derivatives with respect to canonically normalized fields.    The transformation between $x$ and the canonical normalized field $\tau_1^c$ is $\frac{dx}{d\tau_1^c} =  -\sqrt{3/2}/M_P\,x +
\eO(\lambda\,x\,\tau^{3/2},\,\xi\,x)$  \cite{Simon+06}.  This includes the transformation between $x$ and $\tau_1$ as well as the standard non-canonical kinetic term of $\tau_1$ (see, e.g., \cite{GKP}).

Slow-roll inflation, giving a nearly scale-invariant spectrum, occurs when $\epsilon_V, \eta_V, \xi_V \ll 1$.   To first order (sufficiently accurate in the slow roll regime), the observables are (e.g., \cite{LiddleLyth00}):
\begin{eqnarray}
r &\simeq& 16\,\epsilon_{\rm V}\,;\,\,\,n_s \simeq 1-6\,\epsilon_V+2\,\eta_V\,,\\
\frac{dn_s}{d \ln k} &\simeq& 16\,\epsilon_{\rm V}\,\eta_{\rm V}-24\,\epsilon_{\rm V}^2-2\,\xi_{\rm V}\\ 
\bigtriangleup^2_R &\simeq& \frac{V_{\text{tot}}/M^4_P}{24 \pi^2 \epsilon_V}\,,
\end{eqnarray}
where $r$ is the tensor to scalar ratio, $n_s$ is the slope of the
spectrum of scalar primordial fluctuations, $\frac{dn_s}{d \ln k}$ is
the ``running'' of the spectrum, i.e. its deviation from a power law
as a function of scale, and $\Delta^2_R$
is the amplitude of the curvature perturbation spectrum. 

In terms of the variables
\begin{eqnarray}
 \!\!\! z_1\! &=& 3\,\frac{(\lambda\,x_0\,\tau_0^{3/2})\,|W_0|^2g_s^4M_P^4}{\gamma\,4 \pi}\,\frac{1-a\tau_0/g_s}{1-4a\tau_0/g_s}\,, \\
 \!\!\! z_2\! &=& \frac{1-a\tau_0/g_s}{1-4a\tau_0/g_s}\,z_1
 ~~~;~~~
 \!\!\! z_3\! = \frac{1}{8}\frac{(\xi\,x_0)\,|W_0|^2g_s^4 M_P^4}{\gamma\,4\pi} 
\end{eqnarray}
the potential at $(x,\tau_0)$ is
\begin{equation}
V=\gamma x \left[x (1-2z_1) + 4x_0 z_2 + 3 z_3x_0^{-1} x^2 \right]
\end{equation}
and the three slow-roll parameters are
\begin{eqnarray}
 \!\! \epsilon_V\!\! &=& \frac{3}{4}\,\left(\frac{2-4z_1+4z_2 + 9z_3}{1-2z_1+4z_2+3z_3}\right)^2\,,  \label{eq:lin1}\\
\!\!  \eta_V \!\!&=& \frac{3}{2}\,\frac{4-8z_1+4z_2+27z_3}{1-2z_1+4z_2+3z_3}\,, \label{eq:lin2} \\
\!\!  \xi_V \!\!&=&\!\!\!\frac{9}{4}\frac{(8{-}16z_1{+}4z_2{+}81z_3)(2{-}4z_1{+}4z_2{+}9z_3)}{(1-2z_1+4z_2+3z_3)^2}\,. \label{eq:lin3}
\end{eqnarray}
The first two equations can then be inverted giving
\begin{eqnarray}
z_1&=&\frac{1}{2}+\frac{2(9\sqrt{3}+2\sqrt{3}\eta - 24 s \sqrt{\epsilon})}{(9\sqrt{3}+\sqrt{3}\eta- 15 s \sqrt{\epsilon}}z_2\\ 
z_3&=&\frac{4}{3}\frac{3\sqrt{3}+\sqrt{3}\eta - 9 s \sqrt{\epsilon}}{9\sqrt{3}+\sqrt{3}\eta-15 s \sqrt{\epsilon}}z_2\,,
\label{eq:zs1}
\end{eqnarray}
but  the third parameter is given by the other two 
\begin{equation}
\xi_V = - 6(3\sqrt{3}+2\sqrt{3}\eta_V-11 s \sqrt{\epsilon_V}) s \sqrt{\epsilon_V} \, .
\label{eq:consistency1}
\end{equation}
Here $s$ stands for a sign correlated with the direction of rolling:  $s < 0$ when $\CV$ decreases and $s > 0$ when it increases. 

This dependence between $\epsilon_V,\eta_V,\xi_V$ implies that we can pick  $\{n_s,\,r\}$, which are linear in the slow roll parameters, as independent observables and $dn_s/d\ln k$ is determined from them.  Eq. ~(\ref{eq:consistency1}) gives a consistency relation for the running of the spectral index:
\begin{footnotesize}
$$
 \frac{dn_s}{d\log k} =  \nonumber
$$
\begin{equation}
 \frac{24\sqrt{3}\,(2+n_s)\,\sqrt{r} - 9\sqrt{3}\,r^{3/2} + 2\,(25+8n_s)\,sr + 3\,sr^2}{16s}\,.
 \label{eq:run1}
\end{equation}
\end{footnotesize}
For $r\ll 1$ the sign of the running tells us the sign of the direction of rolling of the field. In terms of $\{n_s,r \}$, 
\begin{eqnarray}
 \!\!\! z_1\!&=&\! \!\frac{136\sqrt{3} + 8\sqrt{3}\,n_s + 3\sqrt{3}\,r + 60 \,\sqrt{r}}{6 Q}(4\frac{a\tau_0}{g_s}-1) \label{eq:inv1} \\
 \!\!\! z_3\!\! &=&\! \! \frac{2(40\sqrt{3} + 8\sqrt{3}\,n_s + 3\sqrt{3}\,r + 36\,\sqrt{r})}{9Q} (\frac{a\tau_0}{g_s}-1)\,. \label{eq:inv2}
\end{eqnarray}
with
$Q=(8\sqrt{3}\,n_s + 3\sqrt{3}\,r + 4 \,\sqrt{r}\,(11+4 a\tau_0/g_s) + 8\sqrt{3}\,(5+12a\tau_0/g_s))$.
These equations relate the cosmological observables to the topological and geometric parameters. 

There are two possible branches to roll. Let $x_f$ denote the value of
$x$ where inflation ends and $z=x_f/x_0 = 1+\delta$, where $|\delta|
\ll 1$ and can be positive or negative. Where $\delta > 0$, $\CV$
decreases and where $\delta < 0$, $\CV$ increases. To compute the
number of e-foldings, we must determine where inflation ends.  In our
model this happens because the force in the direction of the second
field, $\tau$, becomes larger than the force along $x$. The steepness
of the potential along the $\tau$ direction means that after this
point, the rolling becomes kinetic energy--dominated and inflation
stops.  It is shown in \cite{Simon+06} that if inflation ends in this
way via fast roll in $\tau$, then $n_s \lap 1$.

The number of e-foldings is:
\begin{eqnarray}
 N_e &=& \int_{t_i}^{t_{e}} Hdt \simeq \frac{1}{M^2_{\rm P}} \int^{\phi_i}_{\phi_{e}} \frac{V}{V'} d\phi.
\end{eqnarray}
Solving this integral we obtain a relation between cosmological parameters and number of efoldings \cite{Simon+06},
\begin{equation}
e^{-9N_e} \approx r.
\label{eq:nefold}
\end{equation}
Thus $r$ is undetectable if we impose the currently favored number of
e-foldings, i.e. $55-70$.  Along with the consistency relation
(Eq.\ref{eq:run1}), this implies that $dn_s/d\ln k$ is also vanishingly small, but its sign depends on the direction of rolling. While this is allowed by current WMAP
\cite{Spergeletal06} data $ -0.11 \le dn_s/d\ln k \le 0.0098$ at 95\%
confidence, other data combinations would give slightly different
conclusions (see table 1). Forthcoming observations (e.g. ACT; Planck
\footnote{www.hep.upenn.edu/act/ ,
www.rssd.esa.int/Planck} ) will improve
the measurement precision by about an order of magnitude. Since we find that $r$ is exponentially suppressed, any  
measurement of a non-zero running will eliminate our models.
\begin{table}
\begin{tabular}{c|ccc}
data& min  & max & in/out \\
set & $d n_s/d\ln k$ &   $d n_s/d\ln k$ & at 95\%CL \\
\hline
WMAP   &-0.116 &  0.0098&   in\\
WMAP+Bolometers &  -0.112 & 0.0027 & in \\
WMAP+HEMP  &-0.12 & -0.00807 & out \\
WMAP+SDSS  & -0.109 & -0.0066 & out\\
WMAP+2dFGRS &-0.11 &0.0027 & in \\
\end{tabular}
\label{table:run}
\caption{95\% confidence region for the running of the spectral index for various data sets combinations. WMAP+Bolometers means WMAP three-year data combined with Boomerang and ACBAR, WMAP+HEMP means WMAP three-year data with CBI and VSA. This model is allowed for some data sets combinations, but is disfavored by other. Ranges have been obtained from the publicly  available  LCDM+running model Markov Chains on LAMBDA \cite{lambda}.}
\end{table}

{\em Measuring the topology}.---We can now express the topological and geometric parameters in terms of observables:
\begin{eqnarray}
\gamma&=&\frac{3\pi^2 r \Delta_R^2M_P^4}{8x_0^2} \frac{5+n_s+12y_0}{y_0-1} \label{gammaobs}\\
\lambda \tau_0^{3/2}&=&\frac{\pi^3}{12}\frac{r \Delta_R^2}{x_0^3g_s^4 |W_0|^2}(17+n_s)\left(\frac{4y_0-1}{y_0-1}\right)^2 \label{lambdaeq} \\
\xi&=&\frac{8 \pi^3}{3}\frac{r \Delta_R^2}{x_0^3g_s^4 |W_0|^2}(5+n_s) \label{xieq} \\
Ae^{-\frac{a\tau_0}{g_s}}&=&\frac{\pi^3}{4}\frac{r \Delta_R^2}{x_0^2g_s^4|W_0|}\frac{17+n_s}{y_0}\frac{4y_0-1}{y_0-1} \label{Aeq}
\end{eqnarray}
where $y_0=a\tau_0/g_s$.   Eqs. (\ref{lambdaeq},\ref{xieq}) yield
\begin{equation}
\xi=32 \lambda \tau_0^{3/2}\frac{5+n_s}{17+n_s}\left(\frac{y_0-1}{4y_0-1}\right)^2\sim 1/3 \lambda \tau_0^{3/2}\,.
\end{equation}
In this setting, inflation begins and ends entirely within the large
volume region and thus our analysis is within the regime of validity
of the approximations (\ref{eq:approx}). It is shown in \cite{Simon+06} that if vacuum energy after inflation ends is  
positive (i.e. $V_{min} > 0$), then $1 < y_0 \lap 2$.  Using (\ref{gammaobs}) one shows that $V_{\rm min}\sim M_P^4 \, r \, \Delta_R^2$.
Since the  amplitude of primordial perturbations amplitude is $\Delta_R^2 \sim 10^{-9}$,
(\ref{eq:nefold}) implies that $V_{\rm min}$ is exponentially small in the
number of e-foldings.  Hence it is much smaller than the measured dark
energy density ($\sim 10^{-120} M_P^4$).  The relation between $r$ and
the vacuum energy implies in our setting that phenomenologically
viable inflation automatically leads to a tiny vacuum energy density.
 
However, to evade accelerator bounds on Kaluza-Klein particle masses,
it turns out that we must have $g_s\gtrsim 10^{-13}$, and the internal
space cannot be too large ($x_0 \gtrsim 10^{-19}$) \cite{Simon+06}.
Then for generic values of $W_0$ (typically of $O(1)$
\cite{statspapers}), (\ref{lambdaeq},\ref{xieq}) imply that realizing
inflation requires exponentially small $\lambda$ and $\xi$.  While
there is no phenomenologically impediment to choosing such small
parameters, they cannot arise as topological invariants of a
Calabi-Yau manifold (e.g., the Euler number is an integer!).  Thus,
for $W_0\sim 1$, requiring that our potential both gives rise to
inflation and originate in string theory, eliminates it as models of
nature.

To realize inflation with topological parameters in a reasonable range, we have to pick an exponentially small $W_0^2 \sim r$:
 given
 $\xi>\xi_*=\zeta(3)/(2\pi)^3\simeq 0.0036$, the required range of $W_0$ is \cite{Simon+06}
\begin{equation}
\frac{r\Delta_R^2}{x_0 g_s} \ll |W_0|^2 \ll \frac{16\pi^5}{\xi_*}10^{30}e^{-9 N_e} \, .
\end{equation}
While this kind of slow-roll model is statistically
disfavored \cite{statspapers}, it is not ruled out.
However, even if its parameters are realizable in string theory, if inflation
required either small internal volumes or $g_s > 1$, further
corrections to the potential (Eq. 2-5) would be needed for consistency.  Even if
these checks are passed, there is a further test: given a
Calabi-Yau manifold, the parameters $A$ and $a$ in the potential are
determined by the theory, but must match (\ref{Aeq}).  This provides a
strategy for systematically falsifying stringy inflation models of
this kind.

{\em Conclusions}--- Robust predictions of these models are {\it i)}
the tensor to scalar ratio $r \approx 0$; any future detection of
primordial gravity waves would falsify this model, {\it ii)}
$dn_s/d\ln k \sim 0$;  $n_s \lap 1$; the sign of running is related to the direction of rolling, {\it (iii)} generally, in these models
there are consistency relations between observables, {\it (iv)}
specifically in this model the value of the cosmological constant, $r$
and $dn_s/d\ln k$ are related and exponentially small in the number of
efoldings.  Our results show that constraints can be imposed on the
topology of the hidden dimensions from cosmological observations.
Further, the requirement that the scalar potential arises from a
string theory is a stringent limitation. This approach provides a
strategy for systematically ruling out stringy inflation models.
\acknowledgments
This work is supported by NSF grants PIRE-0507768 (RJ, LV), PHY-0331728 (JS, VB) and PHY-0355074 (PB), NASA grant ADP03-0000-0092 (LV), and DOE  grant DE-FG02-95ER40893 (JS, VB). We acknowledge the use of the Legacy Archive for Microwave Data (LAMBDA). Support for LAMBDA is 
provided by the NASA Office of Space Science. LV thanks the Astrophysics Theory group at Caltech for hospitality.

\bibliographystyle{apsrev}

\begin{thebibliography}{17}
\expandafter\ifx\csname natexlab\endcsname\relax\def\natexlab#1{#1}\fi
\expandafter\ifx\csname bibnamefont\endcsname\relax
  \def\bibnamefont#1{#1}\fi
\expandafter\ifx\csname bibfnamefont\endcsname\relax
  \def\bibfnamefont#1{#1}\fi
\expandafter\ifx\csname citenamefont\endcsname\relax
  \def\citenamefont#1{#1}\fi
\expandafter\ifx\csname url\endcsname\relax
  \def\url#1{\texttt{#1}}\fi
\expandafter\ifx\csname urlprefix\endcsname\relax\def\urlprefix{URL }\fi
\providecommand{\bibinfo}[2]{#2}
\providecommand{\eprint}[2][]{\url{#2}}


\bibitem{inflation}
\bibinfo{author}{\bibfnamefont{A.~H.} \bibnamefont{{Guth}}},
  \bibinfo{journal}{Phys. Rev. D} \textbf{\bibinfo{volume}{23}}, \bibinfo{pages}{347}
  (\bibinfo{year}{1981});~~~
  \bibinfo{author}{\bibfnamefont{V.~F.} \bibnamefont{{Mukhanov}}}
  \bibnamefont{and} \bibinfo{author}{\bibfnamefont{G.~V.}
  \bibnamefont{{Chibisov}}}, \bibinfo{journal}{JETP Letters} \textbf{\bibinfo{volume}{33}},
  \bibinfo{pages}{532} (\bibinfo{year}{1981});~~~
  \bibinfo{author}{\bibfnamefont{K.}~\bibnamefont{{Sato}}},
  \bibinfo{journal}{MNRAS} \textbf{\bibinfo{volume}{195}}, \bibinfo{pages}{467}
  (\bibinfo{year}{1981});~~~
  \bibinfo{author}{\bibfnamefont{A.~H.} \bibnamefont{{Guth}}} \bibnamefont{and}
  \bibinfo{author}{\bibfnamefont{S.-Y.} \bibnamefont{{Pi}}},
  \bibinfo{journal}{Phys. Rev. Lett.} \textbf{\bibinfo{volume}{49}},
  \bibinfo{pages}{1110} (\bibinfo{year}{1982});~~~
  \bibinfo{author}{\bibfnamefont{S.~W.} \bibnamefont{{Hawking}}},
  \bibinfo{journal}{Phys. Lett. B} \textbf{\bibinfo{volume}{115}},
  \bibinfo{pages}{295} (\bibinfo{year}{1982});~~~
  \bibinfo{author}{\bibfnamefont{A.~D.} \bibnamefont{{Linde}}},
  \bibinfo{journal}{Phys. Lett. B} \textbf{\bibinfo{volume}{108}},
  \bibinfo{pages}{389} (\bibinfo{year}{1982});~~~
  \bibinfo{author}{\bibfnamefont{A.~A.} \bibnamefont{{Starobinsky}}},
  \bibinfo{journal}{Phys. Lett. B} \textbf{\bibinfo{volume}{117}},
  \bibinfo{pages}{175} (\bibinfo{year}{1982});~~~
  \bibinfo{author}{\bibfnamefont{A.}~\bibnamefont{{Albrecht}}} \bibnamefont{and}
  \bibinfo{author}{\bibfnamefont{P.~J.} \bibnamefont{{Steinhardt}}},
  \bibinfo{journal}{Phys. Rev. Lett.} \textbf{\bibinfo{volume}{48}},
  \bibinfo{pages}{1220} (\bibinfo{year}{1982}).

\bibitem[{\citenamefont{{Bin{\'e}truy} and {Gaillard}}(1986)}]{stringinflation}
\bibinfo{author}{\bibfnamefont{P.}~\bibnamefont{{Bin{\'e}truy}}}
  \bibnamefont{and} \bibinfo{author}{\bibfnamefont{M.~K.}
  \bibnamefont{{Gaillard}}}, \bibinfo{journal}{\prd}
  \textbf{\bibinfo{volume}{34}}, \bibinfo{pages}{3069} (\bibinfo{year}{1986});~~~
\bibinfo{author}{\bibfnamefont{T.}~\bibnamefont{{Banks}}},
  \bibinfo{author}{\bibfnamefont{M.}~\bibnamefont{{Berkooz}}},
  \bibinfo{author}{\bibfnamefont{S.~H.} \bibnamefont{{Shenker}}},
  \bibinfo{author}{\bibfnamefont{G.}~\bibnamefont{{Moore}}}, \bibnamefont{and}
  \bibinfo{author}{\bibfnamefont{P.~J.} \bibnamefont{{Steinhardt}}},
  \bibinfo{journal}{\prd} \textbf{\bibinfo{volume}{52}}, \bibinfo{pages}{3548}
  (\bibinfo{year}{1995});~~~
\bibinfo{author}{\bibfnamefont{G.}~\bibnamefont{{Dvali}}} \bibnamefont{and}
  \bibinfo{author}{\bibfnamefont{S.-H.~H.} \bibnamefont{{Tye}}},
  \bibinfo{journal}{Phys. Lett. B} \textbf{\bibinfo{volume}{450}},
  \bibinfo{pages}{72} (\bibinfo{year}{1999});~~~
\bibinfo{author}{\bibfnamefont{C.~P.} \bibnamefont{{Burgess}}},
  \bibinfo{author}{\bibfnamefont{M.}~\bibnamefont{{Majumdar}}},
  \bibinfo{author}{\bibfnamefont{D.}~\bibnamefont{{Nolte}}},
  \bibinfo{author}{\bibfnamefont{F.}~\bibnamefont{{Quevedo}}},
  \bibinfo{author}{\bibfnamefont{G.}~\bibnamefont{{Rajesh}}}, \bibnamefont{and}
  \bibinfo{author}{\bibfnamefont{R.-J.} \bibnamefont{{Zhang}}},
  \bibinfo{journal}{JHEP}
  \textbf{\bibinfo{volume}{7}}, \bibinfo{pages}{47} (\bibinfo{year}{2001});~~~
\bibinfo{author}{\bibfnamefont{J.}~\bibnamefont{{Garc{\'{\i}}a-Bellido}}},
  \bibinfo{author}{\bibfnamefont{R.}~\bibnamefont{{Rabad{\'a}n}}},
  \bibnamefont{and} \bibinfo{author}{\bibfnamefont{F.}~\bibnamefont{{Zamora}}},
  \bibinfo{journal}{JHEP}
  \textbf{\bibinfo{volume}{1}}, \bibinfo{pages}{36} (\bibinfo{year}{2002});
  S.~Kachru, R.~Kallosh, A.~Linde, J.~Maldacena, L.~McAllister and S.~P.~Trivedi,
    JCAP {\bf 0310}, 013 (2003).
   
\bibitem[Blanco-Pillado et al.(2004)]{racetrack1} Blanco-Pillado et al., 
JHEP, 11, 63 (2006).

\bibitem{conquevinf}
  J.~P.~Conlon and F.~Quevedo,
  JHEP {\bf 0601}, 146 (2006).
  

\bibitem[{\citenamefont{{Quevedo}}(2002)}]{Quevedo02}
\bibinfo{author}{\bibfnamefont{F.}~\bibnamefont{{Quevedo}}},
  \bibinfo{journal}{Class. Quant. Grav.}
  \textbf{\bibinfo{volume}{19}}, \bibinfo{pages}{5721} (\bibinfo{year}{2002}).
  
  
  
\bibitem{GKP}
  S.~B.~Giddings, S.~Kachru and J.~Polchinski,
  Phys.\ Rev.\ D {\bf 66}, 106006 (2002).
  
\bibitem[{\citenamefont{{Kachru} et~al.}(2001)\citenamefont{{Kachru},
  {Kallosh}, {Linde}, and {Trivedi}}}]{KKLT}
\bibinfo{author}{\bibfnamefont{S.} \bibnamefont{{Kachru}}},
  \bibinfo{author}{\bibfnamefont{R.}~\bibnamefont{{Kallosh}}},
  \bibinfo{author}{\bibfnamefont{A.}~\bibnamefont{{Linde}}},
  \bibnamefont{and}
  \bibinfo{author}{\bibfnamefont{S. P.} \bibnamefont{{Trivedi}}},
  \bibinfo{journal}{Phys. Rev. D}
  \textbf{\bibinfo{volume}{68}}, \bibinfo{pages}{046005} (\bibinfo{year}{2003}).
  
  
\bibitem[Balasubramanian et al.(2005)]{VijayPerQuevedo} 
  Balasubramanian, V., \& Berglund, P.\ 2004, JHEP, 
11, 85;~~~  Balasubramanian, V., Berglund, P., Conlon, J.~P., \& Quevedo, F.\ 2005, 
JHEP, 3, 7

  
  
\bibitem{CQS}
  J.~P.~Conlon, F.~Quevedo and K.~Suruliz,
  JHEP {\bf 0508}, 007 (2005);~~~
  B.~C.~Allanach, F.~Quevedo and K.~Suruliz,
  arXiv:hep-ph/0512081.
  
  
  
\bibitem{BBHL}
  K.~Becker, M.~Becker, M.~Haack and J.~Louis,
  JHEP {\bf 0206}, 060 (2002).
  
\bibitem{BKQ}
  C.~P.~Burgess, R.~Kallosh and F.~Quevedo,
  JHEP {\bf 0310}, 056 (2003).

\bibitem[{\citenamefont{{Simon} et~al.}(2006)\citenamefont{{Simon}, {Jimenez},
  {Verde}, {Berglund}, and {Balasubramanian}}}]{Simon+06}
\bibinfo{author}{\bibfnamefont{J.}~\bibnamefont{{Simon}}},
  \bibinfo{author}{\bibfnamefont{R.}~\bibnamefont{{Jimenez}}},
  \bibinfo{author}{\bibfnamefont{L.}~\bibnamefont{{Verde}}},
  \bibinfo{author}{\bibfnamefont{P.}~\bibnamefont{{Berglund}}},
  \bibnamefont{and}
  \bibinfo{author}{\bibfnamefont{V.}~\bibnamefont{{Balasubramanian}}},
  \eprint{arXiv:astro-ph}.

\bibitem[{\citenamefont{{Spergel} et~al.}(2006)\citenamefont{{Spergel et~al.}}}]{Spergeletal06}
\bibinfo{author}{\bibfnamefont{D.}~\bibnamefont{{Spergel et~al.}}},
  \eprint{arXiv:astro-ph/0603449}.

\bibitem[{\citenamefont{{Liddle} and {Lyth}}(1992)}]{LiddleLyth92}
\bibinfo{author}{\bibfnamefont{A.~R.} \bibnamefont{{Liddle}}} \bibnamefont{and}
  \bibinfo{author}{\bibfnamefont{D.~H.} \bibnamefont{{Lyth}}},
  \bibinfo{journal}{Phys. Lett. B} \textbf{\bibinfo{volume}{291}},
  \bibinfo{pages}{391} (\bibinfo{year}{1992}).

  
\bibitem{LiddleLyth00}
A.R.~Liddle and    D.H.~Lyth, {\it Cosmological inflation and large-scale structure}, p.~414, Cambridge Press, Cambridge, U.K, 2000.

\bibitem{statspapers}
  F.~Denef and M.~R.~Douglas,
  JHEP {\bf 0405}, 072 (2004).

\bibitem{lambda}
http://lambda.gsfc.nasa.gov/

\end{thebibliography}

\end{document}